\documentclass{aa}
\usepackage{color}
\usepackage{graphicx}
\usepackage{txfonts}
\usepackage[colorlinks, citecolor=blue]{hyperref}
\usepackage{mathtools}
\usepackage{ulem}
\usepackage[dvipsnames]{xcolor}
\usepackage{natbib}
\bibpunct{(}{)}{;}{a}{}{,}

\begin{document}

   \title{Gaia DR2 Gravitational Lens Systems I: \\New lensed quasar candidates around known quasars}
   \author{A. Krone-Martins\inst{1},
   		   L. Delchambre\inst{2},
           O. Wertz\inst{3},
           C. Ducourant\inst{4},
           F. Mignard\inst{5},
           R. Teixeira\inst{6},
           J. Kl\"{u}ter\inst{7},
           J.-F. Le Campion\inst{4},
           L. Galluccio\inst{5},
           J. Surdej\inst{2},
		   U. Bastian\inst{7},
           J. Wambsganss\inst{7},
           M.J. Graham\inst{8},
           S.G. Djorgovski\inst{8},
           E. Slezak\inst{5}
        }

  \institute{
  		 CENTRA, Faculdade de Ci\^encias, Universidade de Lisboa, 1749-016 Lisboa, Portugal\\
          \email{algol@sim.ul.pt}
          \and
         Institut d'Astrophysique et de G\'{e}ophysique, Universit\'{e} de Li\`{e}ge, 19c, All\'{e}e du 6 Ao\^{u}t, B-4000 Li\`{e}ge, Belgium
         \and
         Argelander-Institut f\"{ u}r Astronomie, Universit\"{ a}t Bonn,  Auf dem H\"{ u}gel 71, 53121 Bonn, Germany
          \and
          Laboratoire d'Astrophysique de Bordeaux, Univ. Bordeaux, CNRS, B18N, all{\'e}e Geoffroy Saint-Hilaire, 33615 Pessac, France
          \and
         Universit\'{e} C\^{o}te d'Azur, Observatoire de la C\^{o}te d'Azur, CNRS, Laboratoire Lagrange, Boulevard de l'Observatoire, CS 34229, 06304 Nice, France
         \and
         Instituto de Astronomia, Geof\'isica e Ci\^encias Atmosf\'ericas, Universidade de S\~{a}o Paulo, Rua do Mat\~{a}o, 1226, Cidade Universit\'aria, 05508-900 S\~{a}o Paulo, SP, Brazil
         \and
        Zentrum f\"{u}r Astronomie der Universit\"{a}t Heidelberg, Astronomisches Rechen-Institut, M\"{o}nchhofstr. 12-14, 69120 Heidelberg, \\Germany
         \and
         California Institute of Technology, 1200 E. California Blvd, Pasadena, CA 91125, USA
          }

   \date{Received April 30, 2018; accepted ???, ???}

  \abstract
   {Strong gravitationally lensed quasars are among the most interesting and useful observable extragalactic phenomena. Because their study constitutes a unique tool in various fields of astronomy, they are highly sought, not without difficulty. Indeed, even in this era of all-sky surveys, their recognition remains a great challenge, with barely a few hundred currently known systems.}
   {In this work we aim to detect new strongly lensed quasar candidates in the recently published {\it Gaia} Data Release 2 (DR2), which is the highest spatial resolution astrometric and photometric all-sky survey, attaining effective resolutions from 0.4" to 2.2".}
   {We cross-matched a merged list of quasars and candidates with the {\it Gaia} DR2 and found 1,839,143 counterparts within 0.5". We then searched matches with more than two {\it Gaia} DR2 counterparts within 6". We further narrowed the resulting list using astrometry and photometry compatibility criteria between the {\it Gaia} DR2 counterparts. A supervised machine learning method, Extremely Randomized Trees, is finally adopted to assign to each remaining system a probability of being lensed.}
   {We report the discovery of three quadruply-imaged quasar candidates that are fully detected in {\it Gaia} DR2. These are the most promising new quasar lens candidates from {\it Gaia} DR2 and a simple singular isothermal ellipsoid lens model is able to reproduce their image positions to within $\sim$1 mas. This letter demonstrates the gravitational lens discovery potential of {\it Gaia}.}
   {}
\titlerunning{{\it Gaia} DR2 Gravitational Lenses I: New lensed quasars candidates}
\authorrunning{A. Krone-Martins et al.}
\maketitle

\section{Introduction}
Strong gravitational lenses \citep[hereafter GL]{1936Sci....84..506E, 1937PhRv...51..290Z} probe many key features in astronomy: e.g. dark matter halos of galaxies \citep[e.g.][]{2004ApJ...610..663O}, substructures in lensing galaxies \citep[e.g.][]{2002ApJ...567L...5M}, the determination of the Hubble constant independently of the cosmic distance ladder \citep[e.g.][]{1964MNRAS.128..307R, 2017MNRAS.468.2590S}, properties of dark energy \citep[e.g.][]{2015ApJ...806..185C}, and they might even reveal the shape of the central engine of quasars \citep[e.g.][]{2018MNRAS.475.1925T}. 
However, since the first detection of a multiply-imaged quasar \citep[][]{1979Natur.279..381W} they have been elusive. The discovery and characterization of these systems require exceptional imaging capabilities, posing a challenge to present all-sky surveys from the ground. Thus, the limited number of reliable lensed systems has historically plagued many of the potential studies that can be performed with these objects, mainly due to systematics at the level of individual lenses. 

The data from the ESA/{\it Gaia} space mission \citep{2016A&A...595A...1G} is expected to change this situation dramatically. At the present time {\it Gaia} is conducting the largest, most precise and accurate all-sky astrometric survey from space. Its main goal is to produce a three-dimensional dynamical map of the Milky Way based on the measurement of positions, parallaxes, proper motions supplemented by spectro-photometric parameters for more than $10^9$ stars. Moreover, the instruments also detect and transmit to the ground observations of millions of galaxies and quasars \citep{2012A&A...543A.100R, 2013A&A...556A.102K, 2014A&A...568A.124D, 2015A&A...576A..74D, 2016A&A...595A...1G}. Thus a careful analysis of the {\it Gaia} Data Releases presents a unique opportunity to perform the first all-sky magnitude-limited census of strongly lensed quasars, down to image separations as small as $\sim 0.18$".

Recently, \citet{2016A&A...590A..42F} have shown that, from the $\sim 6.6 \times 10^5$ quasars that would likely be observed by the satellite, nearly 2900 could be multiply-imaged and resolved in the final {\it Gaia} Data Release.
From these, about  2650 are expected to be double-imaged lenses and  250 systems would consist of more than two lensed images. Therefore the survey may lead to an increase in the number of lensed quasars by more than an order of magnitude with respect to what is currently known. Thus, {\it Gaia} will provide invaluable data on new lensed systems, which will be studied either individually or from a statistical point of view.
 
This letter is the first published report from a group specially set up to unravel the aforementioned possibilities of {\it Gaia}, the {\it Gaia Gravitational Lenses}, or {\it Gaia GraL}. We aim to perform a systematic investigation of gravitational lenses in the various {\it Gaia} Data Releases (DRs), through lens candidate detection from {\it Gaia} data, ground-based observations and physical modeling of the lenses. We report here our analysis of the environment of 1,839,143 quasar candidates compiled from the literature, with a positional counterpart in the {\it Gaia} Data Release 2 \citep[DR2,][]{GaiaDR2MainPaper}.
Based on these data coupled with statistical cuts and machine learning methods, we discover three new quadruply-imaged quasar candidates.

In Sect.\ref{Data} we present the {\it Gaia} DR2 data adopted in this letter and the compilation of a quasar list. The lensed quasar candidate selection criteria and the probability assignment method are detailed in Sect.\ref{Method}. We comment on a few candidates and discuss our results in Sect.\ref{Discussion}. Finally, we summarize our findings and briefly present our future developments in Sect.\ref{Conclusions}. 
 
\section{Data}\label{Data}
\subsection{The Gaia data}\label{GaiaData}
Using the {\it Gaia} archive facility at ESAC \citep{2017A&C....21...22S}, we extracted the positions ($\alpha, \delta$), astrometric pseudo-colors \citep{GaiaDR2AstrometricSolution}, parallaxes ($\varpi$), proper-motions ($\mu_\alpha, \mu_\delta$), fluxes in the $G$, $G_{BP}$ and $G_{RP}$ passbands \citep{GaiaDR2Photometry}, together with their associated uncertainties, from the table {\tt gaiadr2.gaia\_source} via an ADQL positional cross-match with the compiled quasar list (Sec. \ref{QuasarCatalog}).

Although the {\it Gaia} on-board instrument can attain a spatial resolution of $\sim0.18$", the {\it Gaia} DR1 \citep[GDR1;][]{2016A&A...595A...2G} reached an effective angular resolution that was limited to the range $2$" -- $4$"
\citep{2017A&A...599A..50A}. Indeed, the raw data processing and the final filtering based on the  astrometric quality prevented a large fraction of objects with small separations to be present in early {\it Gaia} Data Releases, as these objects are prone to larger errors. However, because the astrometric quality improves at each release the expected angular resolution found in the releases gets better with the updated processing.

In {\it Gaia} DR2, the effective resolution now reaches $\sim 0.4$", with completeness for separations larger than $\sim 2.2$" \citep{GaiaDR2MainPaper}. Moreover, due to the instrument design, the evolution of the data processing and calibration, the effective angular resolution is not the same for different types of {\it Gaia} DR2 data. The aforementioned resolution applies strictly only for astrometry and G band photometry. The spatial resolution of {\it Gaia} spectrophotometry in {\it Gaia} DR2 is $\sim 2$", while its completeness reaches $\sim 3.5$" \citep{GaiaDR2MainPaper}.

\subsection{The compiled quasar list}\label{QuasarCatalog}
We first set up a large list of quasars by compiling and merging several catalogs of quasars or quasar candidates. Since these catalogs were constructed using different selection criteria based on spectroscopy, photometry, cuts in visible or near-infrared bands, the resulting list is inhomogeneous in reliability and quality of prior information.

The major contribution to our list comes from the largest presently available catalog, the Million Quasars Catalog \citep[MILLIQUAS;][]{2015PASA...32...10F,Flesch(2017)}, that comprises 1,998,464 quasars or high-confidence quasar candidates over the entire sky. Further, we considered the ALLWISE catalog of quasars \citep[1,354,775 objects, ][]{Secrest(2015)}. To avoid creating duplicates, we search for MILLIQUAS sources within 1"; when a counterpart is not found, we add the ALLWISE source to our compiled quasar list. The same procedure is applied to the LQAC3 \citep[321,945 sources, ][]{Souchay(2015)} and the SDSS Quasar catalog DR12Q \citep[297,301 sources, ][]{Paris(2017)}. The final list contains 3,112,975 objects, whose 1,839,143 have a counterpart in the {\it Gaia} DR2  within an angular separation smaller than 0.5". 

As this list is intended to be exhaustive, it might contain contaminants. To discard the most obvious ones, we first filter out the Galactic plane (we only kept $|\, b \, | > 15\degr$) and the Magellanic clouds area. Then we apply a soft astrometric test to exclude objects characterized by $\varpi -3 \sigma_\varpi > 4$ mas and $|\, \mu\, | + 3 \sigma_{\mu} > 4$ mas/yr, where $\mu$ stands for $\mu_{\alpha^{*}}$ and $\mu_{\delta}$. The reasons that motivate the choice of these thresholds are based on  the {\it Gaia} DR2 properties of known lenses; more details regarding these thresholds are given in Paper II (Ducourant et al. in prep.). This step eliminates about $1\%$ of the candidates in the quasar list.
\section{Method}
\label{Method}

\subsection{Neighbors extraction and parameter cuts: clusters}\label{Neighbors}
We matched the {\it Gaia} DR2 with detections within a 6" radius in the direction of each object of our quasar list, and kept record of quasars with one or more neighbors. The quasar neighbors were also astrometrically filtered as described in Sect. \ref{QuasarCatalog}, eliminating $\sim 10\%$ of them. The resulting systems are called {\it clusters}. Afterwards, we applied a series of tests between the {\it Gaia} DR2 detections to verify $3\sigma$ statistical compatibility between their astrometric ($\varpi$, $\mu_{\alpha^{*}}$, $\mu_\delta$ and astrometric pseudo-color) and photometric parameters (colors computed from $G$, $G_{BP}$ and $G_{RP}$ fluxes when available in {\it Gaia} DR2). These tests aimed to discard remaining stellar contaminants, and might also discard the lensing galaxy (if detected). The tests, for instance $\mid \varpi_i - \varpi_j \mid \le 3 \sigma_{\mid \varpi_i - \varpi_j \mid}$ for the parallax, were applied between each possible pair $(i, j)$ of the {\it Gaia} DR2 sources in each cluster. The clusters with at least one pair of {\it Gaia} DR2 sources passing the tests were considered for further analysis, as described in Sec. \ref{ertprob}.

Considering the tests applied on astrometric data, 16,500 clusters have two {\it Gaia} DR2 sources, 1,874 clusters have three, 269 have four and 66 have five or more {\it Gaia} DR2 sources. This less stringent test is adopted to define the clusters for which the lensing probability is computed (Sec. \ref{ertprob}). However, for a subset of these sources, the {\it Gaia} DR2 colors can be computed. The number of clusters that passed {\it all} the tests is 372; split in 8, 46 and 320 characterized respectively by four, three and two {\it Gaia} DR2 sources, defining more reliable candidate clusters. These results are summarized in Table \ref{tab:clusters}. We note that from the doubly-imaged systems that passed {\it all} the tests, 121 clusters have separations smaller than 3", while 28 clusters attain 1". Only one of these 28 stringently selected clusters is a previously known gravitational lens, SDSS1332+0347.

\begin{table}[!ht]
\begin{center}
\caption{\label{tab:clusters} The number of candidate clusters that passed the tests. Note that many sources cannot be tested for color in the {\it Gaia} DR2.}
\begin{tabular}{lrrr}
\hline
\hline
{\it Gaia} DR2		&    Astrometry &	Astrometry & Astrometry\\
sources	    &               &	 \& Color  &	 \& Color, $\Delta\theta\le1"$\\
\hline
2             		&    16,500 &	320 &	28\\
3 					&     1,874 &	46  &	0\\
4 				    &       269 &	8   & 0\\
$\ge$ 5 			&        66 &	0   & 0\\
\hline
\end{tabular}
\end{center}
\end{table}

Although several catalogs used in our compiled quasar list explicitly exclude known GLs, our list contains counterparts for 136 known GLs. In this set there are 21 quadruply-imaged systems, out of which 19 have two or more {\it Gaia} DR2 counterparts, and 9 have four counterparts. From 11 previously known triply-imaged systems, 10 were detected with two or more counterparts, while from the 104 doubly-imaged systems, 79 show two counterparts. Thus from the 136 known GLs in our quasar list, we found 108 GLs with two or more {\it Gaia} DR2 counterparts.

\subsection{Identification of lens candidates from clusters}\label{ertprob}
The next step consists in classifying the identified clusters with respect to their chance of being a GL candidate. A GL candidate is defined here as a configuration of images (i.e. relative positions and differences in magnitudes) that can be well-reproduced with a non-singular isothermal ellipsoid in presence of an external shear model \citep[NSIEg,][]{1994A&A...284..285K,1987ApJ...312...22K}. This model effectively describes the mass distribution of massive early-type galaxies in the region where multiple images occur \citep{2017arXiv171204945G}. The complete procedure we adopted to discriminate GL candidates from fortuitous groups of stars is detailed in Paper III (Delchambre et al. in prep.), which is dedicated to a large blind-search for gravitational lenses in the entire {\it Gaia} DR2 catalog. In this section, we provide the reader with an overall description of the basic principles.
    
During the process, we assign to each candidate a probability based on Extremely Randomized Trees \citep[hereafter ERT]{geurts2006}. This probability reflects the match between a candidate and the learning set used to build the ERT and, thus, does not constitute a probability in the mathematical sense. These can however be matched to an expected ratio of identification of the GL candidates (the true positive rate, hereafter TPR) and to an expected ratio of misclassification of groups of stars as GL candidates (the false positive rate, hereafter FPR) through the use of a cross-validation procedure. Our ERT models were trained using a set of 106,623,188 simulated NSIEg lensed image configurations composed of more than three components, uniformly covering the most common range of parameters of the lens model. An equal number of contaminant observations were also produced to correspond to fortuitous groups of stars in our learning sample. However, as these simulations do not follow a realistic distribution of galaxies/stars parameters, the derived TPR and FPR are estimates only. Each of these simulated configurations was then altered by the addition of noise, in agreement with uncertainties of the {\it Gaia} astrometric and photometric observations. Also, as we expect some lensed images to be missing from the {\it Gaia} DR2, all combinations of the three and four images were considered for building the ERT. We note that the fifth image from the NSIEg lens model, being often out of reach of the {\it Gaia} photometric sensitivity, was accordingly not taken into account. These ERT models will be referred to in the following as ABCD, ABC, ABD, ACD and BCD where A, B, C, D identify the images we used during the learning phase of the corresponding ERT, assuming these are sorted in ascending order of $G$ magnitude. 
    
    All the 3- and 4-image configurations were then processed using the aforementioned models. The TPR associated with a probability above $0.9$ regarding the models ABCD, ABC, ABD, ACD, BCD are respectively given by $0.741$, $0.398$, $0.391$, $0.244$ and $0.274$, while the corresponding FPR are given by $0.006$, $0.025$, $0.024$, $0.022$ and $0.024$.

\section{The new lensed quasar candidates in {\it Gaia} DR2}
\label{Discussion}
We report the first results of this effort in searching GL among the current and upcoming {\it Gaia} data releases. We discovered three new candidates, namely GRAL113100-441959, GRAL203802-400815, and GRAL122629-454209, each of them being composed of four lensed images. 

The first two candidates are characterized by an ERT probability higher than \(0.95\). Beside those two newly discovered candidates, five known lensed systems were also re-discovered with similar probabilities: HE0435-1223, PG1115+080, PKS1830-211, RXJ1131-1231, and WFI2033-4723. The lens candidate GRAL122629-454209 is characterized by a lower ERT probability, due to large flux uncertainties. This comes as no surprise since the sources that compose this system are close to the {\it Gaia} detection limit. At this regime the {\it Gaia} DR2 photometry might show significant biases. Moreover, this faint candidate also suffers from being at a nearby distance ($\sim$2") to a bright (G$\sim$16.3) source, and some of their observations might even share a single assigned window \citep[e.g.][]{GaiaDR2Photometry}.

\begin{figure*}[!htp]
\begin{center}
\includegraphics[width=0.245\textwidth]{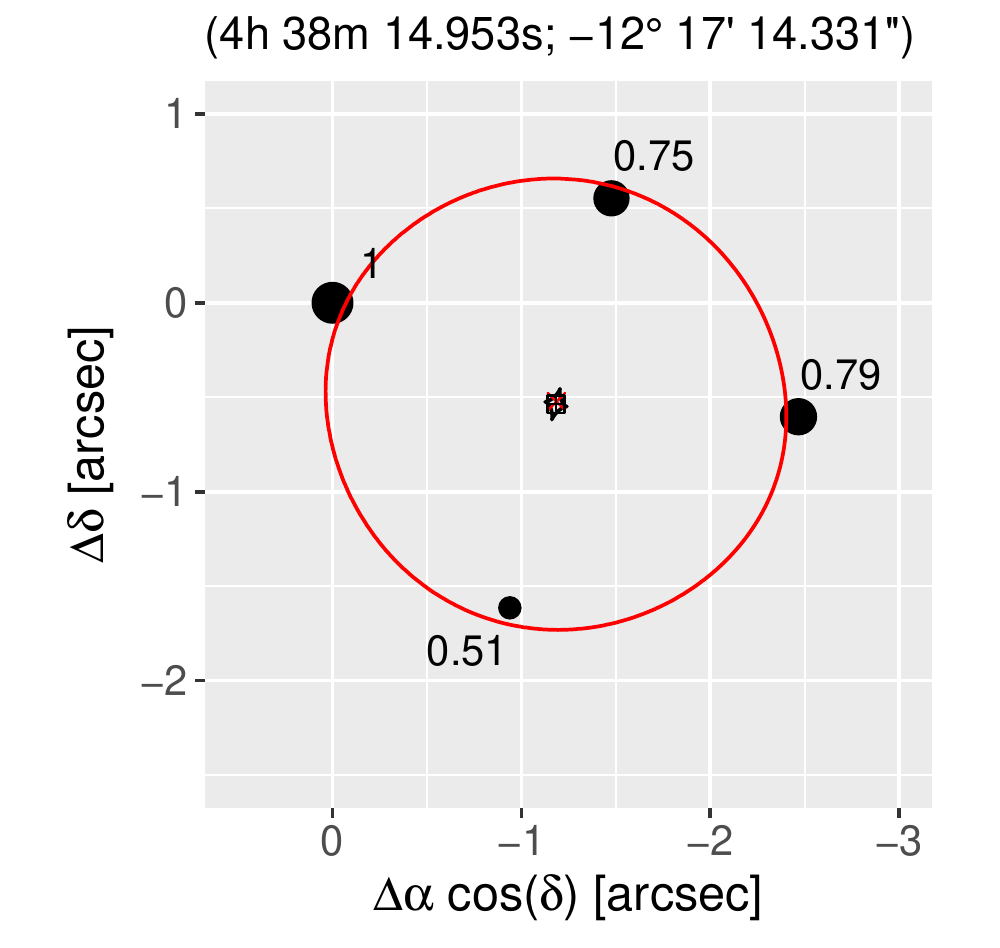}
\includegraphics[width=0.245\textwidth]{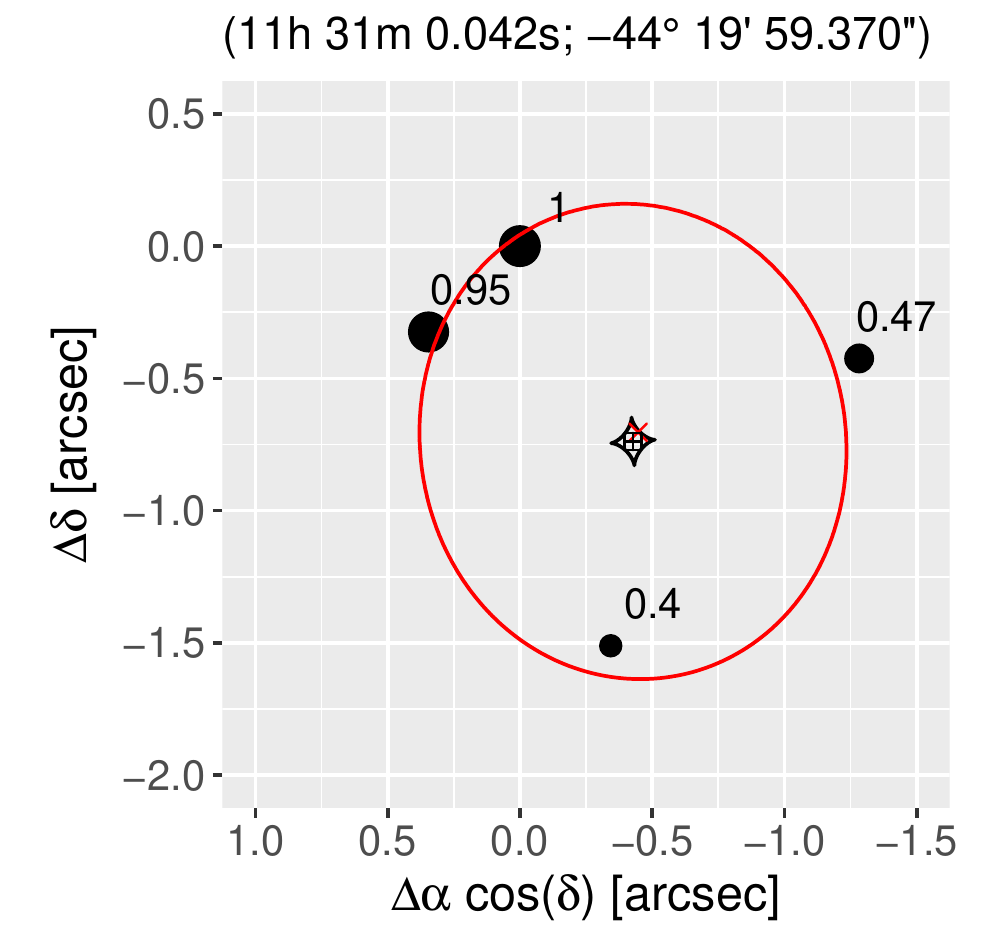}
\includegraphics[width=0.245\textwidth]{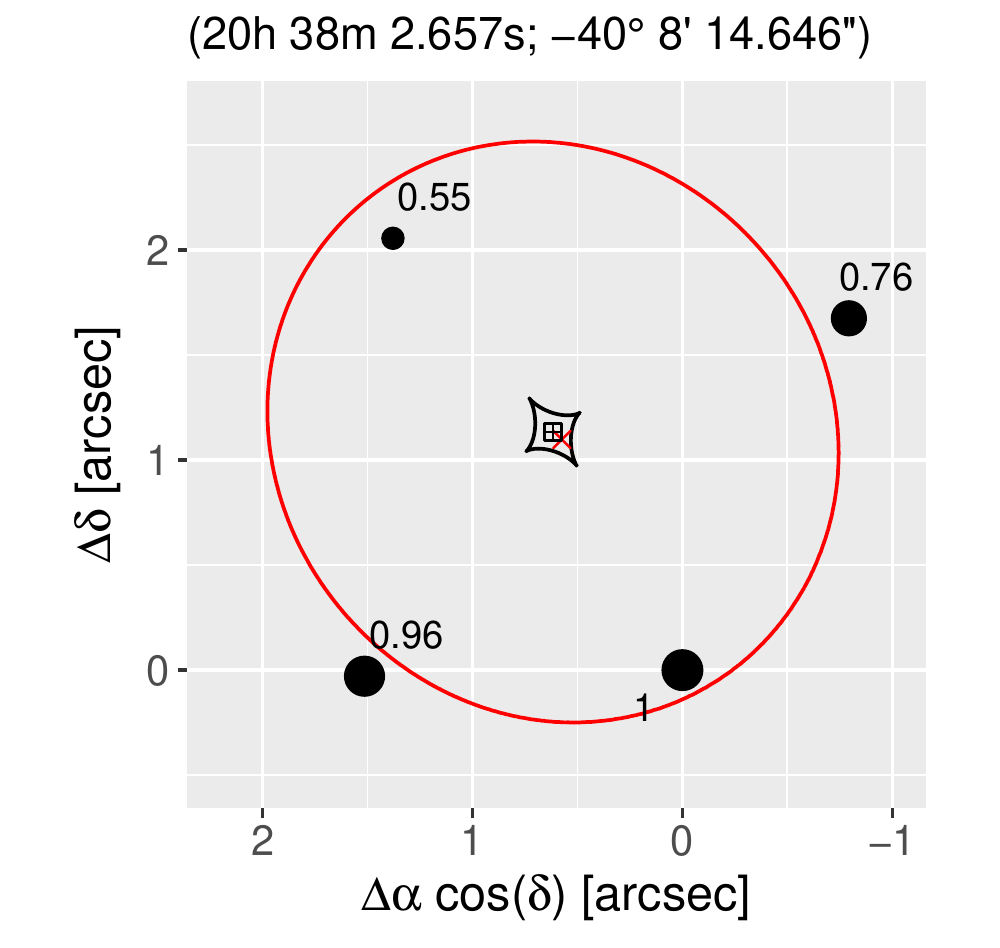}
\includegraphics[width=0.245\textwidth]{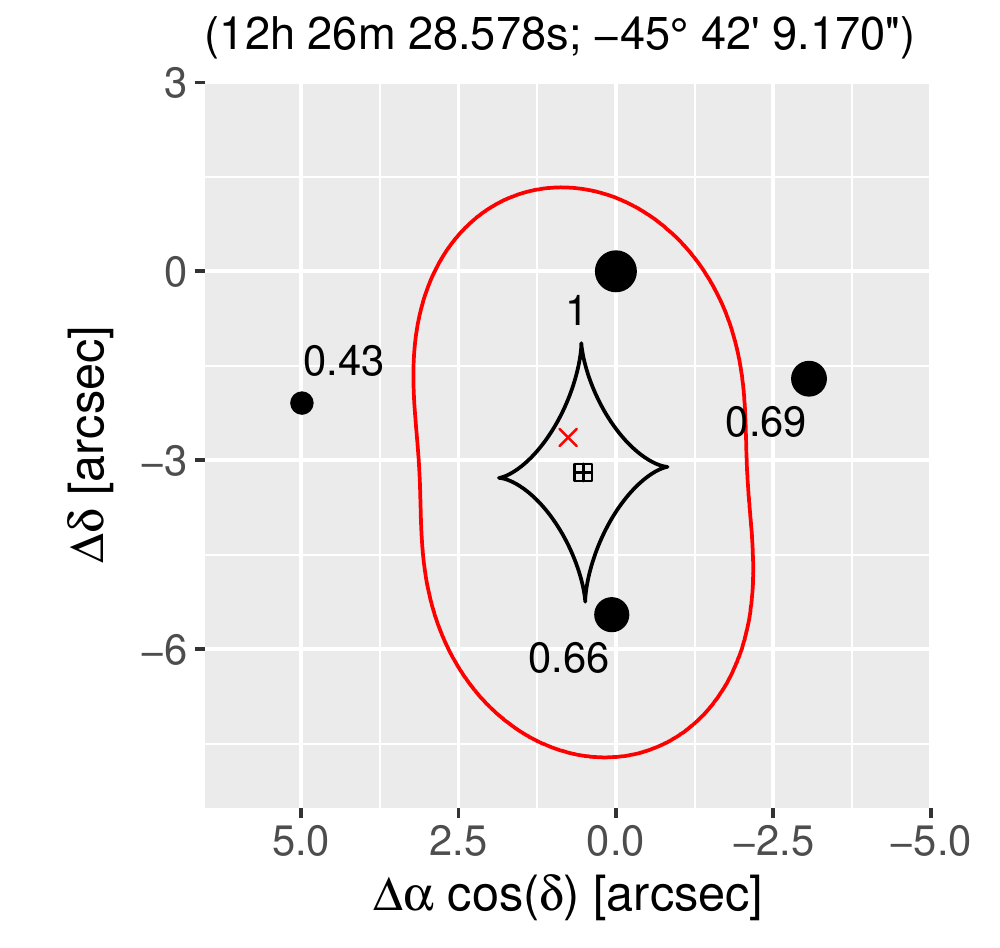}
\caption{Examples of strong gravitationally lensed quasars and candidates detected from {\it Gaia} DR2. Black dots locate the astrometric positions from {\it Gaia} DR2. The numbers associated with each position indicate the G band flux ratios with respect to the brightest image. The leftmost chart corresponds to the well-known lens HE0435-1223. The following three systems are newly discovered quadruply-imaged candidates from {\it Gaia} DR2 data alone. Several products obtained from our simple lens modeling are represented: the critical curves (red), the caustics (black), the source positions (red crosses), and the predicted deflector centroids (black squares).
\label{fig:examplelenscandidates}}
\end{center}
\end{figure*}

As a first step in the validation process, the three new systems were modeled using the public code {\tt lensmodel} \citep[v1.99, ][]{Keeton_computationalMethods_2001}. We adopted a simple SIE model plus external shear, using the lensed image positions and G-band flux ratios, except for GRAL122629-454209, as observational constraints. The image positions are extremely well-reproduced (to within 1 mas) whereas the flux ratios are in reasonable agreement with those measured. A more detailed Bayesian modeling of the {\it Gaia GraL} quasar lens candidates is presently in preparation (Wertz et al. in prep.).
The {\it Gaia} relative astrometry and photometry of the new quasar lens candidates are reported in Table \ref{tab:gralrelative} and illustrated in Fig.~\ref{fig:examplelenscandidates}, which also shows some modeling results. 

In addition to the candidates reported here, we also found $65$ systems composed of three detections that are characterized by an ERT probability of being a quadruply-imaged system higher than \(0.95\). These candidates, along with systems characterized with lower probabilities and doubly-imaged candidates, will be subject of dedicated studies.

\begin{table*}[!ht]
\begin{center}
\caption{\label{tab:gralrelative} Relative astrometry and flux ratios of the three newly discovered strong gravitationally lensed quasar candidates from {\it Gaia} DR2. The astrometric errors are typically smaller than $\sim$ 1 mas for all sources. The larger flux ratio errors in the A\&B images of the GRAL113100-441959 candidate might originate from their smaller separation ($\sim0.4"$), at the limit of the effective resolution of {\it Gaia} DR2.}
\begin{tabular}{l|rrr|rrr|rrr}
\hline
\hline
            &      \multicolumn{3}{c|}{    GRAL113100-441959} &      \multicolumn{3}{c|}{    GRAL203802-400815} &      \multicolumn{3}{c}{    GRAL122629-454209} \\
\hline
Image		&    $\Delta\alpha\cos(\delta)$ &	$\Delta\delta$ & G-band&    $\Delta\alpha\cos(\delta)$ &	$\Delta\delta$ & G-band&    $\Delta\alpha\cos(\delta)$ &	$\Delta\delta$ & G-band\\
		    &    ['']           &	['']           & flux ratio&    ['']           &	['']           & flux ratio&    ['']           &	['']           & flux ratio\\
\hline
A           &     0.000(2) &  0.000(1) &	1.000(17) &    0.000(1) &  0.000(1) &	1.000(9) &    0.000(1) &  0.000(1) &	1.000(6)\\
B           &     0.345(1) & -0.325(1) &	0.947(10) &    1.515(1) & -0.029(1) &	0.963(5) &   -3.066(1) & -1.705(1) &	0.688(3)\\
C           &    -1.282(1) &  0.425(1) &	0.475(6) &   -0.793(1) &  1.676(1) &	0.763(4) &    0.065(1) & -5.453(1) &	0.659(3)\\
D           &    -0.343(1) & -1.511(1) &	0.402(5) &    1.379(2) &  2.057(2) &	0.550(8) &    4.986(2) & -2.092(2) &	0.434(4)\\
\hline
\end{tabular}
\end{center}
\end{table*}

\section{Conclusions}
\label{Conclusions}
In this letter we report the discovery of three new gravitationally lensed quasar candidates directly extracted from {\it Gaia} Data Release 2. This result demonstrates the great potential of the ESA/{\it Gaia} mission in this field, and that the series of {\it Gaia} Data Releases will provide an invaluable dataset particularly well-suited for extragalactic and cosmological studies.

This study is the first {\it Gaia}-data only detection of lensed quasar candidates. We have shown that from a state-of-the art list of quasars and quasar candidates comprising more than 3.1 million objects, 1.8 million of them have a counterpart in the {\it Gaia} DR2. By applying astrometric and photometric selection criteria, we have derived a list of {\it Gaia} DR2 source clusters and using a machine learning method trained with a NSIEg lens model, we have derived lensing probabilities for the system candidates composed of three and four components. 
Finally, we have modeled a selection of promising lens candidates, demonstrating that they are sufficiently reliable for further investigations. Of course, the lensing nature of the systems presented in this letter will require further spectroscopic and high-angular resolution multi-band imaging observations to be confirmed. 

{\it Gaia} was not particularly designed for extragalactic studies. However, it is a transversal all-sky survey that will touch many expected and unexpected areas of astronomy, and this letter is a vivid illustration of this capability. {\it Gaia} promises substantial advances in the study of strong lensed, multiply-imaged quasars as, by its final Data Release, it will provide the highest spatial resolution of any all-sky survey. Moreover, it also provides exquisite astrometric and photometric data of the observed sources, including time-series in the final Data Releases. {\it Gaia} will thus enable to set strong constraints in the modeling of gravitationally lensed systems, and on cosmological parameters derived from one of the most interesting phenomena in nature.

\begin{acknowledgements}
       AKM acknowledges the support from the Portuguese Funda\c c\~ao para a Ci\^encia e a Tecnologia (FCT) through grants SFRH/BPD/74697/2010, from the Portuguese Strategic Programme UID/FIS/00099/2013 for CENTRA, from the ESA contract AO/1-7836/14/NL/HB and from the Caltech Division of Physics, Mathematics and Astronomy for hosting a research leave during 2017-2018, when this paper was prepared. 
      LD acknowledges support from the ESA PRODEX Programme `{\it Gaia}-DPAC QSOs' and from the Belgian Federal Science Policy Office.
      OW is supported by the Humboldt Research Fellowship for Postdoctoral Researchers.
      SGD and MJG acknowledge a partial support from the NSF grants AST-1413600 and AST-1518308, and the NASA grant 16-ADAP16-0232.
      We acknowledge partial support from `Actions sur projet INSU-PNGRAM', and from the Brazil-France exchange programmes Funda\c c\~ao de Amparo \`a Pesquisa do Estado de S\~ao Paulo (FAPESP) and Coordena\c c\~ao de Aperfei\c coamento de Pessoal de N\'ivel Superior (CAPES) -- Comit\'e Fran\c cais d'\'Evaluation de la Coop\'eration Universitaire et Scientifique avec le Br\'esil (COFECUB).
      This work has made use of the computing facilities of the Laboratory of Astroinformatics (IAG/USP, NAT/Unicsul), whose purchase was made possible by the Brazilian agency FAPESP (grant 2009/54006-4) and the INCT-A, and we thank the entire LAi team, specially Carlos Paladini, Ulisses Manzo Castello, Luis Ricardo Manrique and Alex Carciofi for the support.
      This work has made use of results from the ESA space mission {\it Gaia}, the data from which were processed by the {\it Gaia} Data Processing and Analysis Consortium (DPAC). Funding for the DPAC has been provided by national institutions, in particular the institutions participating in the {\it Gaia} Multilateral Agreement. The {\it Gaia} mission website is:
http://www.cosmos.esa.int/gaia. Some of the authors are members of the {\it Gaia} Data Processing and Analysis Consortium (DPAC).
\end{acknowledgements}

\bibliographystyle{aa}
\bibliography{bibliography} 
\end{document}